# Stationary Charge Radiation in Anisotropic Photonic Time Crystals

Huanan Li,[1] Shixiong Yin,[2,3] Huan He,[1] Jingjun Xu,[1,*] Andrea Alù,[2,3,4,†] and Boris Shapiro[5,‡]

[1]MOE Key Laboratory of Weak-Light Nonlinear Photonics, School of Physics, Nankai University,

Tianjin 300071, China

[2]Photonics Initiative, Advanced Science Research Center, City University of New York,

New York, New York 10031, USA

[3]Department of Electrical Engineering, City College of The City University of New York,

New York, New York 10031, USA

[4]Physics Program, Graduate Center, City University of New York, New York,

New York 10016, USA

[5]Department of Physics, Technion-Israel Institute of Technology, Haifa 32000, Israel

*Time metamaterials exhibit a great potential for wave manipulation, drawing increasing attention in recent years. Here, we explore the exotic wave dynamics of an anisotropic photonic time crystal (APTC), formed by an anisotropic medium whose optical properties are uniformly and periodically changed in time. Based on a temporal transfer matrix formalism, we show that a stationary charge embedded in an APTC emits radiation, in contrast to the case of isotropic photonic time crystals, and its distribution in momentum space is controlled by the APTC band structure. Our approach extends the functionalities of time metamaterials, offering new opportunities for radiation generation and control, with implications for both classical and quantum applications.*

[*] jjxu@nankai.edu.cn  
[†] aalu@gc.cuny.edu  
[‡] boris@physics.technion.ac.il

The electrodynamics of space-time metamaterials, i.e., artificial materials whose parameters are modulated in space and time, has become in recent years a very active field of research [1]-[7]. In the special case of pure-time media [2], a wave propagating in a spatially homogeneous medium changes its propagation features as the material properties change uniformly in time [8]-[11]. The simplest example is a plane wave subject to a sudden change of permittivity $\varepsilon$ and/or permeability $\mu$. This problem was studied in the last century by Morgenthaler [12], who showed that the wave, with given frequency $\omega$ and wave vector $\vec{k}$, splits into forward and backward propagating waves. Both waves retain the original value of $\vec{k}$ due to preserved spatial uniformity of the medium, but oscillate at a new frequency corresponding to the new dispersion of the modified medium. An insightful discussion of this time-reflection effect, with its experimental demonstration for both water and electromagnetic waves, appears in [13]-[15]. In such a context, a variety of different platforms for temporal modulation have been considered in the last few years, enabling intriguing phenomena and/or mechanisms, such as the inverse prism [16], temporal aiming [17], complete polarization conversion [18], extreme energy transformations [19], temporal disorder [20]-[22] and nonreciprocity [23].

Of great interest are also problems concerning radiation by sources (charges, electric dipoles, quantum emitters) embedded in time-varying media [24]-[30]. Already in the 70's it has been pointed out by Ginzburg and co-workers that a charge moving with constant velocity in a homogeneous medium is bound to radiate if the refractive index changes in time [24]-[27]. This early work focused mostly on a single switching event of the material properties. Recently, it has been shown that photonic time crystals (PTCs), created by periodically modulating the scalar permittivity $\varepsilon(t)$ of a homogeneous isotropic material [31]-[32], can induce amplified coherent radiation, associated with momentum-gap modes, of both free moving electrons [28] and oscillating dipoles [30], based on which the interesting concept of non-resonant PTC lasers has been proposed.

So far, PTCs have been limited to isotropic media. In this paper, we introduce anisotropic photonic time crystals (APTCs), in which the periodic temporal modulation pertains to the (relative) permittivity tensor $\bar{\bar{\varepsilon}}(t)$ characterizing uniform anisotropic media. We construct a generalized temporal transfer matrix formalism to deal with sources embedded in APTCs, showing that the anisotropy plays a fundamental role in enriching the radiation features of these time metamaterials. In particular, we show that a stationary charge embedded in an APTCs emits radiation, in stark contrast with light emission of free electrons in PTCs, for which charge motion is required [28]. Based on our results, we unveil an implicit relation between light emission of the fixed charge and the band structure of the APTC, demonstrating exciting opportunities to control the radiative energy distribution in momentum space for steering the direction of light emission in space.

*Anisotropic photonic time crystals and their band structure* — We start with the discussion of APTCs without sources, focusing on the simplest form of APTC. Specifically, our APTC is formed by alternating periodically in time a pair of lossless non-magnetic media, one being isotropic and the other one being a uniaxial crystal [labeled by $(m)$, $m = 1, 2$ respectively] [see Fig. **1**(a)]. For simplicity, we neglect material dispersion and use the principal axes of the uniaxial crystal as our coordinate system. In these axes, the dielectric tensor of the uniaxial crystal is $\bar{\bar{\varepsilon}}^{(2)} = \text{diag}(\varepsilon_\perp, \varepsilon_\perp, \varepsilon_\parallel)$, while the isotropic crystal is described by $\bar{\bar{\varepsilon}}^{(1)} = \varepsilon \bar{\bar{I}}_3$ with $\bar{\bar{I}}_N$ being the identity matrix of order $N$. In the APTC, the displacement field $\vec{D}^{(r)}(\vec{r}, t)$ of the total radiation can be written in terms of the space Fourier transform as $\vec{D}^{(r)}(\vec{r}, t) = \frac{1}{(2\pi)^3} \int_{k_z>0} \vec{D}^{(r)}(\vec{k}, t) e^{-j\vec{k}\cdot\vec{r}} d^3\vec{k} + c.c.$, where the wave vector $\vec{k} = [k_x, k_y, k_z]^T$, with $c.c.$

standing for complex conjugation, and superscripts $(r)$ and $T$ denoting radiation waves and the transpose operation, respectively. Within each temporal interval in which the medium is uniaxial, for given wave vector $\vec{k}$, the dispersion relation allows for two different *positive* frequencies $\omega_1^{(2)} = kc_0/\sqrt{\varepsilon_\perp}$ and $\omega_2^{(2)} = kc_0\sqrt{(\hat{k}_x^2 + \hat{k}_y^2)/\varepsilon_\parallel + \hat{k}_z^2/\varepsilon_\perp}$, where $c_0$ is the speed of light in free space, and $\hat{k}_\alpha \equiv k_\alpha/k$, $\alpha = x, y, z$ [with wave number $k \equiv |\vec{k}|$] are the direction cosines of $\vec{k}$. The two frequencies correspond to two waves, i.e., the ordinary and extraordinary waves, propagating in the $\vec{k}$-direction (forward) and having mutually orthogonal electric displacements, $\vec{D}_1 = [-\hat{k}_y, \hat{k}_x, 0]^T$ and $\vec{D}_2 = [\hat{k}_x\hat{k}_z, \hat{k}_y\hat{k}_z, -(\hat{k}_x^2 + \hat{k}_y^2)]^T$. These two vectors, $\vec{D}_1(\vec{k})$ and $\vec{D}_2(\vec{k})$, together with the vector $\vec{k}$, form an orthogonal triplet. In each temporal interval in which the medium is isotropic, the pair $\vec{D}_1$ and $\vec{D}_2$ can be chosen as a basis supporting two independent forward waves with the same positive frequency $\omega_1^{(1)} = \omega_2^{(1)} = kc_0/\sqrt{\varepsilon}$, essentially treating the medium as a degenerate uniaxial crystal.

Considering the additional two backward waves ($-\vec{k}$-direction) oscillating at the corresponding *negative* frequencies $-\omega_{1,2}^{(m)}$ within each temporal slab of the isotropic ($m = 1$) or the uniaxial ($m = 2$) crystal, the total displacement of the radiation field in momentum space $\vec{D}^{(r)}(\vec{k}, t)$ for the APTC [see Fig. **1**(a)] can be written compactly as

$$\vec{D}^{(r)}(\vec{k}, t) = \sum_{n=1}^{2} \vec{D}_n(\vec{k})[1, \quad 1]\psi_n(t), \tag{1}$$

where the vector $\psi_n(t) \equiv [f_n(t), \quad b_n(t)]^T$ consists of the forward [and backward] time-dependent components $f_n(t) \propto e^{j\omega_n^{(m)}t}$ [and $b_n(t) \propto e^{-j\omega_n^{(m)}t}$] of the first ($n = 1$) and the second ($n = 2$) wave in medium $m$.

We proceed to describe the evolution of the radiation in the APTC, which boils down to studying the time dependence of the total state vector $\psi(t) \equiv \begin{bmatrix} \psi_1(t) \\ \psi_2(t) \end{bmatrix}$ of the two waves in $\vec{k}$ space, since momentum is conserved upon each temporal switching event. Free wave propagation in each temporal slab of the medium $m = 1, 2$ in the APTC is described by the block-diagonal matrix

$$F^{(m)} = \mathrm{diag}\{F_1^{(m)}, \ F_2^{(m)}\} \tag{2}$$

with blocks $F_n^{(m)} = \mathrm{diag}\{e^{j\omega_n^{(m)} T_m}, \ e^{-j\omega_n^{(m)} T_m}\}, n = 1, 2$, which describes the temporal evolution of the state $\psi(t)$ within the slab of duration $T_m$. At a switching event of the APTC, say at switching time $t_s = 0$ [see Fig. **1**(a)], as the medium changes abruptly from $m = 1$ to 2, the state vector $\psi(t)$ experiences an instantaneous variation $\psi(t_s^+) = J^{(2,1)} \psi(t_s^-)$ with matching matrix [33]

$$J^{(2,1)} = \begin{bmatrix} J_1^{(2,1)} & 0 \\ 0 & J_2^{(2,1)} \end{bmatrix}, \ J_n^{(2,1)} \equiv \frac{1}{2\omega_n^{(2)}} \begin{bmatrix} \omega_n^{(2)} + \omega_n^{(1)}, & \omega_n^{(2)} - \omega_n^{(1)} \\ \omega_n^{(2)} - \omega_n^{(1)}, & \omega_n^{(2)} + \omega_n^{(1)} \end{bmatrix}, \tag{3}$$

following from the continuity of $\vec{D}$ and its time derivative $\partial \vec{D}/\partial t$ (or equivalently the magnetic-flux density $\vec{B}$) [5], at the time interface $t_s$. Similarly, we obtain the matching matrix $J^{(1,2)} = [J^{(2,1)}]^{-1}$, accounting for the rapid evolution of the state $\psi(t)$ at the other switching event, say at time $t_s = T_2$ [see Fig. **1**(a)], when the medium changes from $m = 2$ to 1. To study the evolution of $\psi(t)$ in the APTC, the matching matrix $J^{(2,1)}$ in Eq. (3) and the propagation matrix $F^{(m)}$ in Eq. (2) constitute building blocks.

Of primary interest for our purpose is to study the transfer matrix $M \equiv F^{(1)}[J^{(2,1)}]^{-1}F^{(2)}J^{(2,1)}$, which describes the evolution of the state $\psi(t)$ across one temporal unit cell of the APTC, of duration $T = T_1 + T_2$ [an example of such cell is indicated in Fig. **1**(a) by the red dashed box, and $\psi(3T^-) = M\psi(2T^-)$ correspondingly]. This matrix $M$ fully describes the long-time dynamics of the APTC, since the transfer matrix for $N$ periods is simply given by $M^N$. Using Eqs. (2-3), the matrix $M$ turns out to be block-diagonal, i.e., $M = \text{diag}\{M_1, M_2\}$, with each block being $M_n = F_n^{(1)} \left(J_n^{(2,1)}\right)^{-1} F_n^{(2)} J_n^{(2,1)}$, $n = 1, 2$. The first and the second block $M_1$ and $M_2$ are associated with the ordinary and the extraordinary waves of the uniaxial crystal, respectively. Using the blocks $M_n$, we obtain their band structure separately by solving the secular equations

$$\det\{M_n(\vec{k}) - e^{j\Omega_{n\sigma}T}\bar{\bar{I}}_2\} = 0, n = 1, 2, \qquad (4)$$

where $\Omega_{n\sigma}$ is the Floquet frequency, with the sub-index $\sigma = \pm$ labeling different eigenvalues. In Fig. **1**(b-c), we consider the APTC shown schematically in Fig. **1**(a) with parameters $\varepsilon = 1$, $\varepsilon_\perp = 25$, $\varepsilon_\parallel = 4$ and $T_1 = T_2 = T/2$ [38], and plot the complex dispersion relation of the ordinary [Fig. **1**(b)] and extraordinary [Fig. **1**(c)] waves propagating in the $k_x - k_z$ plane. To understand these band structures, we note that the blocks $M_n(\vec{k}), n = 1,2$ in Eq. (4) can be parametrized as $M_n = \begin{bmatrix} a_n & d_n \\ d_n^* & a_n^* \end{bmatrix}$, and satisfy $\det M_n = 1$ based on Eqs. (2-3) (see also [21]). Correspondingly, their eigenvalues are given by $\lambda_{n\pm} \equiv e^{j\Omega_{n\pm}T} = \text{Re}\, a_n \pm \sqrt{(\text{Re}\, a_n)^2 - 1}$. If $(\text{Re}\, a_n)^2 < 1$, we have $|\lambda_{n+}| = 1/|\lambda_{n-}| = 1$ and thus $\text{Im}(\Omega_{n\pm}T) = 0$, leading to oscillations in time without systematic growth or decay. On the other hand, when $(\text{Re}\, a_n)^2 > 1$, the eigenvalues $\lambda_{n\pm}$ are real and $|\lambda_{n+}| = 1/|\lambda_{n-}| \neq 1$. In this case, $\text{Im}(\Omega_{n\pm}T) \neq 0$ and one eigenstate grows exponentially in time while the other decays. Generally, which of these two cases is realized depends on $\vec{k}$, producing bands and gaps in momentum space.

For ordinary waves (block $M_1$), the band structure is consistent with the one of PTCs, and it does not depend on the propagation direction of the waves [see Fig. **1**(b)], following from the direction cosines $\hat{k}_\alpha$-independent $\omega_1^{(m)}$ for $M_1$. In contrast, in the case of extraordinary waves, the band structure, determined by the block $M_2$, is anisotropic, since $\Omega_{2\sigma}$ depends not only on the magnitude of the wavevector $\vec{k}$ but also on its direction [see the direction dependence of $\omega_2^{(2)}$ and Fig. **1**(c)]. This feature implies remarkable features unique to APTCs: for instance, whether the waves in APTCs are amplified (and how strongly) depends on their momentum directions. Interestingly, the gaps in the APTC band structure can close when fixing the wave number $k$ [for instance, at $k = 2.5\,k_0$ with $k_0 = 2\pi/Tc_0$ in Fig. **1**(c)] and varying the direction cosines $\hat{k}_\alpha$, or more precisely only the polar angle $\theta$, since, due to the rotational symmetry of the considered APTC, the band structure is invariant with respect to the azimuthal angle $\varphi$.

***Radiation by a stationary charge in APTCs*** — Next, we study the effect of a stationary point charge $Q$ in the considered APTCs. Clearly, the existence of a stationary charge does not affect the free propagation of radiation fields in static media, so that the evolution of the state $\psi(t)$, dictating the radiation $\vec{D}^{(r)}(\vec{k}, t)$ in Eq. (1), is determined by the same propagation matrix $F^{(m)}$ in Eq. (2) within each temporal slab of the APTC [see Fig. **1**(a)]. Nevertheless, due to the presence of the charge $Q$, the effects of the temporal interfaces on the radiation fields can change dramatically. Assume for instance that the charge $Q$ is fixed at the origin $\vec{r} = 0$. As a result, the total displacement field in momentum space becomes $\vec{D}(\vec{k}, t) = \vec{D}_0(\vec{k}, t) + \vec{D}^{(r)}(\vec{k}, t)$, which includes also a particular solution due to the source $Q$

$$\vec{D}_0(\vec{k}, t) = \frac{jQ}{\vec{k} \cdot \bar{\bar{\varepsilon}}(t)\vec{k}} \bar{\bar{\varepsilon}}(t)\vec{k} \tag{5}$$

with the matrix $\bar{\bar{\varepsilon}}(t) = \bar{\bar{\varepsilon}}^{(m)}$, yielding electrostatic parts $\vec{D}_0(\vec{k}, t) = \vec{D}_0^{(m)}(\vec{k})$ respectively, for time $t$ within the temporal slab of the crystal $m = 1$ or 2 [see Fig. **1**(a)]. Correspondingly, based on the same temporal boundary conditions for Eq. (3), the matching relation for the state $\psi(t)$ of the radiated fields becomes [33]

$$\psi(t_s^+) = S^{(2,1)} + J^{(2,1)}\psi(t_s^-) \qquad (6)$$

with the extra source term $S^{(2,1)} = s_{21}[0, \ 0, \ 1, \ 1]^T$ [where the prefactor $s_{21} = \frac{jQ\Delta_\varepsilon \cos\theta}{k(1+\Delta_\varepsilon \cos(2\theta))}$ with $\Delta_\varepsilon \equiv \frac{\varepsilon_\parallel - \varepsilon_\perp}{\varepsilon_\parallel + \varepsilon_\perp}$] for the temporal interface at time $t_s$ when the medium switches from $m = 1$ to 2. Indeed, the source term $S^{(2,1)}$ appears in Eq. (6) due to the mismatch between the electrostatic solutions $\vec{D}_0^{(1)}(\vec{k})$ and $\vec{D}_0^{(2)}(\vec{k})$ of the charge $Q$ in $m = 1$ and 2 [see also Ref. [39]], producing spontaneous emission of *extraordinary* waves even when the radiation fields vanish before switching, i.e., $\psi(t_s^-) = 0$ [25]. We note that the source term $S^{(2,1)} = 0$ when $\varepsilon_\parallel = \varepsilon_\perp$, which corresponds to a conventional (isotropic) photonic time crystal. Indeed, stationary charges are not expected to radiate in such a scenario. Similarly, an extra source term $S^{(1,2)}$, related to $S^{(2,1)}$ in Eq. (6) by $S^{(1,2)} = -[J^{(2,1)}]^{-1}S^{(2,1)} = -S^{(2,1)}$, emerges at the other switching event of the APTC in Fig. **1**(a) when the medium switches from $m = 2$ to 1.

Now, we are ready to study the radiation by a stationary charge $Q$ embedded in the APTC. We assume that initially, i.e., prior to time $t = 0$, there is no radiation field, so that the initial state of radiation $\psi(0^-) = 0$. Then we study the effects of the first modulation cycle of the APTC. At the first switching event at time $t = 0$, the medium switches from the crystal $m = 1$ to 2, producing a radiation field $\psi(0^+) = S^{(2,1)}$ [see Eq. (6)]. This field freely propagates in the medium $m = 2$, for a time interval $T_2$, resulting in $\psi(T_2^-) = F^{(2)}S^{(2,1)}$. Next comes the switching event back to the medium $m = 1$ at time $t = T_2$, which multiples $\psi(T_2^-)$ by the matching matrix $J^{(1,2)}$, and, in addition, it creates a new radiation field $S^{(1,2)}$. Thus, we have $\psi(T_2^+) = S^{(1,2)} + J^{(1,2)}\psi(T_2^-)$. The cycle is completed by free propagation for duration $T_1$ in medium $m = 1$, so that $\psi(T^-) = F^{(1)}\psi(T_2^+)$. Putting these pieces together we can write the radiated field at the end of the first modulation cycle as $\psi(T^-) = (F^{(1)} - M)S^{(1,2)}$, or equivalently $\psi_1(T^-) = 0$ and $\psi_2(T^-) = -s_{21}(F_2^{(1)} - M_2)\begin{bmatrix}1\\1\end{bmatrix}$, where $M = F^{(1)}J^{(1,2)}F^{(2)}J^{(2,1)}$ is the same unit-cell transfer matrix as introduced before. Indeed, it is generally true that the radiation field produced by the APTC in Fig. **1**(a), when $\psi(0^-) = 0$, does not involve ordinary waves, i.e., $\psi_1(t > 0) = 0$, due to the fact that the source term creates only extraordinary waves [see Eq. (6)], and that they propagate independently from the ordinary waves [see Eqs. (2-3)]. Repeating the above recursion $n_t$ times, we obtain the total radiation field after $n_t$ modulation cycles [33]

$$\psi_2(n_t T^-) = -s_{21} \frac{\bar{\bar{I}}_2 - M_2^{n_t}}{\bar{\bar{I}}_2 - M_2}\left(F_2^{(1)} - M_2\right)\begin{bmatrix}1\\1\end{bmatrix}, \qquad (7)$$

which connects the radiation field of the APTC to the band structure of extraordinary waves via $M_2$ [see Eq. (4)].

The APTC with a stationary charge not only supports radiation fields, but it also controls their energy distribution in momentum space. To see this, we study the time-dependent evolution of the energy distribution in momentum space of the radiation field based on Eq. (7). At each time $t = n_t T^-$, the fields are present in the isotropic medium $m = 1$, and thus the total electromagnetic energy $W_{em}(t) = \frac{1}{(2\pi)^3}\int_{k_z>0} w_{em}(\vec{k},t) d^3\vec{k}$ with energy density in momentum space $w_{em}(\vec{k},t) = w_{em}^{(s)}(\vec{k},t) + w_{em}^{(r)}(\vec{k},t)$, which involves the constant electrostatic part $w_{em}^{(s)}(\vec{k},t) = Q^2/(\varepsilon_0 \varepsilon k^2)$ and the radiation part $w_{em}^{(r)}(\vec{k},t) = 2\sin^2\theta\,|\psi_2(t)|^2/(\varepsilon_0\varepsilon)$. In Fig. **2**, we choose the same parameters as in Fig. **1**(b-c), and show the normalized radiative energy density $\widehat{w}_{em}^{(r)}(\vec{k},t) \equiv w_{em}^{(r)}(\vec{k},t)/\max_{\vec{k}} w_{em}^{(r)}(\vec{k},t)$ after different modulation cycles. As shown, the radiative energy density $\widehat{w}_{em}^{(r)}(\vec{k},t)$ spreads over a relatively large momentum space near $k = 0$ initially [Fig. **2**(a)] due to the prefactor $s_{21} \propto 1/k$ of the source in Eq. (7). With increasing modulation cycles, the modes in the bandgaps with nonzero imaginary frequency grow exponentially, and start to dominate in momentum space [Fig. **2**(b)]. When the number of modulation cycles becomes larger [Fig. **2**(c-f)], the radiative energy $\widehat{w}_{em}^{(r)}(\vec{k},t)$ increasingly localizes around the momentum with $k \approx 2.5k_0$ and $\theta \approx 0.094$ [33], where the gap modes have the relatively larger imaginary frequency $|\text{Im}(\Omega_{2\sigma}T)|$ (and thus growing rate) [Fig. **1**(c)] within the $\vec{k}$-range of initial excitations [Fig. **2**(a)], forming a highly directional non-resonant laser source tunable via the manipulation of momentum bandgaps [33]. In contrast with the non-resonant tunable laser based on PTCs proposed in [28],[30], the extreme lasing directionality in APTCs is achieved due to their inherent anisotropic band structure, which does not require source manipulation, such as precise control of the movement of free charges for delicate phase matching required to sustain the radiation process [28].

To illustrate better the evolution of the radiation field in momentum space, we study the distinct dynamics of the magnetic-flux density $\vec{B}(\vec{k},t)$ in the bandgap and band, respectively. For the APTC with a stationary charge, $\vec{B}(\vec{k},t)$ is associated only with the radiation, and it reads $\vec{B}(\vec{k},n_tT^-) = \vec{B}_2^{(1)}(\vec{k})[1, -1]\psi_2(n_tT^-)$ [with the basis state $\vec{B}_2^{(1)}(\vec{k}) = \vec{D}_1(\vec{k})\eta_0/\sqrt{\varepsilon}$] after $n_t$ modulation cycles, where $\eta_0$ is the characteristic impedance of free space. In Fig. **3**, we choose two representative momenta $\vec{k} = \vec{k}_l, l = 1,2$ with spherical coordinates $(k,\theta,\varphi) = (2.5k_0, 0.094, 0)$ and $(2.5k_0, 1.477, 0)$ in the bandgap and band [see Fig. **1**(c)], and show the only nonzero component $B_y(\vec{k}_l, t)$ as a function of time calculated both solving Maxwell's equations numerically [solid lines] and using Eq. (7) [empty circles]. As shown, starting from zero initial values without radiation, the gap mode grows exponentially [Fig. **3**(a) and lower inset] and takes over the bounded band mode [Fig. **3**(b)] after a few modulation cycles, steering the direction of light emission in space, see upper insets for individual snapshots at time $t = 6T$ and [33] for full animations.

*Conclusions* — In this Letter, we introduced the concept of APTCs formed by photonic time crystals involving anisotropic media, and developed a generalized temporal transfer matrix formalism to study APTCs with sources. We showed that APTCs can enable radiation of a stationary charge and manage emission features due to their reconfigurable (anisotropic) band structure in momentum space. Our approach greatly extends the concept of non-resonant tunable PTC lasers proposed in [28],[30] to the scenario of stationary charges, and it showcases unique opportunities to leverage time interfaces, anisotropy and electrostatic fields to efficiently produce directional dynamic non-resonant lasing. This concept may be further generalized by incorporating non-Hermitian elements [19] to push its boundaries towards the full control of light emission via time metamaterials without spatial structures, with implications for laser technologies, classical and quantum photonic applications. Furthermore, these concepts may be fruitfully implemented in other physical domains, such as for acoustic waves.

*Acknowledgements* — This work was supported by the Research Startup Funds of Nankai University (Grant No. ZB22000106), by the Office of Naval Research, the Air Force Office of Scientific Research and the Simons Foundation.

# References

[1] A. M. Shaltout, V. M. Shalaev, and M. L. Brongersma, "Spatiotemporal light control with active metasurfaces," Science **364**, eaat3100 (2019).

[2] C. Caloz and Z.-L. Deck-Léger, "Spacetime metamaterials—Part I: General concepts," IEEE Trans. Antennas Propag. **68**, 1569 (2020); "Spacetime metamaterials—Part II: Theory and Applications," IEEE Trans. Antennas Propag. **68**, 1583 (2020).

[3] N. Engheta, "Metamaterials with high degrees of freedom: space, time, and more," Nanophotonics **10**, 639 (2021).

[4] G. Castaldi, V. Pacheco-Peña, M. Moccia, N. Engheta and V. Galdi, "Exploiting space-time duality in the synthesis of impedance transformers via temporal metamaterials," Nanophotonics, **10**, 3687 (2021).

[5] E. Galiffi, R. Tirole, S. Yin, H. Li, S. Vezzoli, P. A. Huidobro, M.G. Silveirinha, R. Sapienza, A. Alù, and J. B. Pendry, "Photonics of time-varying media," Adv. Photon. **4**, 014002 (2022).

[6] S. Yin, E. Galiffi, and A. Alù, "Floquet metamaterials," eLight **2**, 8 (2022).

[7] Z. Hayran, F. Monticone, "Challenging Fundamental Limitations in Electromagnetics with Time-Varying Systems," arXiv: 2205.07142 (2022).

[8] D. Ramaccia, A. Alù, A. Toscano, and F. Bilotti, "Temporal multilayer structures for designing higher-order transfer functions using time-varying metamaterials," Appl. Phys. Lett. **118**, 101901 (2021).

[9] J. Xu, W. Mai, and D. H. Werner, "Generalized temporal transfer matrix method: a systematic approach to solving electromagnetic wave scattering in temporally stratified structures," Nanophotonics **11**, 1309 (2022).

[10] C. Rizza, G. Castaldi, and V. Galdi, "Short-Pulsed Metamaterials," Phys. Rev. Lett. **128**, 257402 (2022).

[11] E. Galiffi, S. Yin, A. Alù, "Tapered Photonic Switching," Nanophotonics **11**, 3575 (2022).

[12] F. Morgenthaler, "Velocity modulation of electromagnetic waves," IRE Trans. Microw. Theory Tech. **6**, 167 (1958).

[13] V. Bacot, M. Labousse, A. Eddi, M. Fink, and E. Fort, "Time reversal and holography with spacetime transformations," Nat. Phys. **12**, 972 (2016).

[14] V. Bacot, G. Durey, A. Eddi, M. Fink, and E. Fort, "Phase-conjugate mirror for water waves driven by the Faraday instability," Proc. Natl. Acad. Sci. U. S. A. **116**, 8809 (2019).

[15] H. Moussa, G. Xu, S. Yin, E. Galiffi, Y. Radi, and A. Alù, "Observation of Temporal Reflections and Broadband Frequency Translations at Photonics Time-Interfaces," arXiv: 2208.07236 (2022).

[16] A. Akbarzadeh, N. Chamanara, and C. Caloz, "Inverse prism based on temporal discontinuity and spatial dispersion," Opt. Lett. **43**, 3297 (2018).

[17] V. Pacheco-Peña and N. Engheta, "Temporal aiming," Light Sci. Appl. **9**, 129 (2020).

[18] J. Xu, W. Mai, and D. H. Werner, "Complete polarization conversion using anisotropic temporal slabs," Opt. Lett. **46**, 1373 (2021).

[19] H. Li, S. Yin, E. Galiffi, and A. Alù, "Temporal Parity-Time Symmetry for Extreme Energy Transformations," Phys. Rev. Lett. **127**, 153903 (2021).

[20] Y. Sharabi, E. Lustig, and M. Segev, "Disordered Photonic Time Crystals," Phys. Rev. Lett. **126**, 163902 (2021).

[21] R. Carminal, H. Chen, R. Pierrat, and B. Shapiro, "Universal Statistics of Waves in a Random Time-Varying Medium," Phys. Rev. Lett. **127**, 094101 (2021).

[22] B. Apffel, S. Wildeman, A. Eddi, and E. Fort, "Experimental Implementation of Wave Propagation in Disordered Time-Varying Media," Phys. Rev. Lett. **128**, 094503 (2022).

[23] H. Li, S. Yin, and A. Alù, "Nonreciprocity and Faraday Rotation at Time Interfaces," Phys. Rev. Lett. **128**, 173901 (2022).

[24] V. L. Ginzburg, V. N. Tsytovich, "On the theory of transition radiation in a nonstationary medium," Sov. Phys. JETP **38**, 65 (1973).

[25] G. M. Maneva, "On the radiation of a fixed charge in a non-stationary medium (in Russian)," Commun. on Phys. P. N. Lebedev Inst. Ac. of. Sc. USSR **2**, 21 (1976).

[26] V. L. Ginzburg, V. N. Tsytovich, "Several problems of the theory of transition radiation and transition scattering," Phys. Rep. **49**, 1 (1979).

[27] V. L. Ginzburg, V. N. Tsytovich, *Transition Radiation and Transition Scattering* (Adam Hilger, Bristol and New York, 1990).

[28] A. Dikopoltsev, Y. Sharabi, M. Lyubarov, Y. Lumer, S. Tsesses, E. Lustig, I. Kaminer, and M. Segev, "Light emission by free electrons in photonic time-crystals," Proc. Natl. Acad. Sci. USA **119**, e2119705119 (2021).

[29] N. V. Budko, "Electromagnetic radiation in a time-varying background medium," Phys. Rev. A **80**, 053817 (2009).

[30] M. Lyubarov, Y. Lumer, A. Dikopoltsev, E. Lustig, Y. Sharabi, M. Segev, "Amplified emission and lasing in photonic time crystals," Science **377**, 425 (2022).

[31] J. R. Zurita-Sánchez, P. Halevi, and J. C. Cervantes-González, "Reflection and transmission of a wave incident on a slab with a time-periodic dielectric function $\varepsilon(t)$," Phys. Rev. A **79**, 053821 (2009).

[32] E. Lustig, Y. Sharabi, and M. Segev, "Topological aspects of photonic time crystals," Optica **5**, 1390 (2018).

[33] See Supplemental Material for derivations of Eqs. (3,6,7), discussion of experimental aspects and localization points of radiative energy in momentum space, and animations of the time evolution associated with Fig. **3**, which includes Refs. [34]-[37].

[34] C. Rizza, D. Dutta, B. Ghosh et al., "Extreme Optical Anisotropy in the Type-II Dirac Semimetal $NiTe_2$ for Applications to Nanophotonics," ACS Appl. Nano Mater. **Article ASAP** (2022).

[35] G. A. Ermolaev, D. V. Grudinin, Y. V. Stebunov et al., "Giant optical anisotropy in transition metal dichalcogenides for next-generation photonics," Nat. Commun. **12**, 854 (2021).

[36] J. K. H. Wong, K. G. Balmain, and G. V. Eleftheriades, "Fields in Planar Anisotropic Transmission-Line Metamaterials," IEEE Trans. Antennas Propag. **54**, 2742 (2006).

[37] Q. Ma, G. D. Bai, H. B. Jing, C. Yang, L. Li, and T. J. Cui, "Smart Metasurface with self-adaptively reprogrammable functions," Light Sci. Appl. **8**, 98 (2019).

[38] The required parameters for the demonstrated physics in the full text are flexible and appear to be reachable using transition-metal dichalcogenides or transmission-line metamaterials [33].

[39] M. J. Mencagli, D. L. Sounas, M. Fink, and N. Engheta, "Static-to-dynamic field conversion with time-varying media," Phys. Rev. B **105**, 144301 (2022).

**Figures**

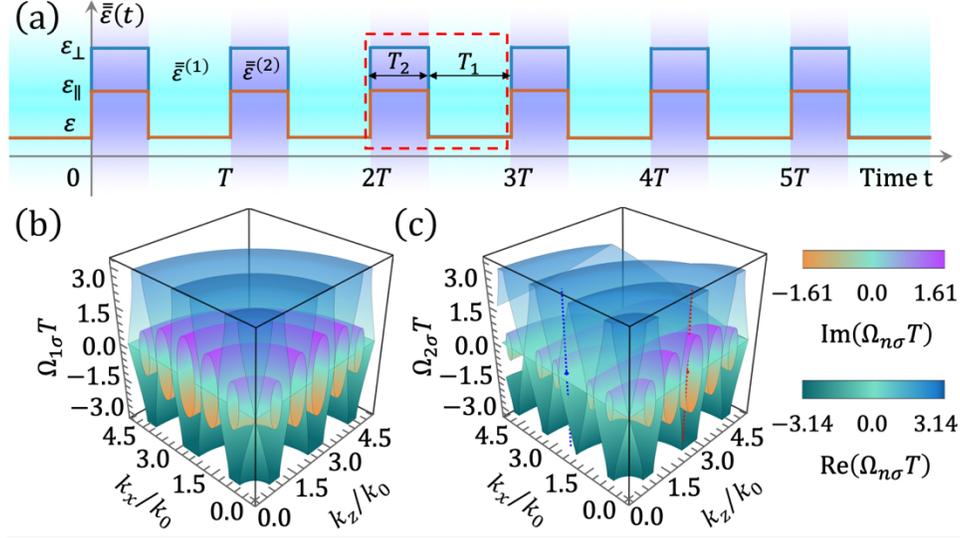

**Fig. 1.** (a) Schematic of the APTC described in the text. (b, c) Band structure of the APTC [with $\varepsilon = 1$, $\varepsilon_\perp = 25$, $\varepsilon_\parallel = 4$ and $T_1 = T_2 = T/2$ in (a)] in momentum $\vec{k}$ (normalized to $k_0 = 2\pi/Tc_0$) for (b) ordinary and (c) extraordinary waves. In (c), the red [and the blue] vertical dashed line intersects the $k_x - k_z$ plane at $\vec{k}$ with the spherical coordinates $(2.5k_0, 0.094, 0)$ [and $(2.5k_0, 1.477, 0)$].

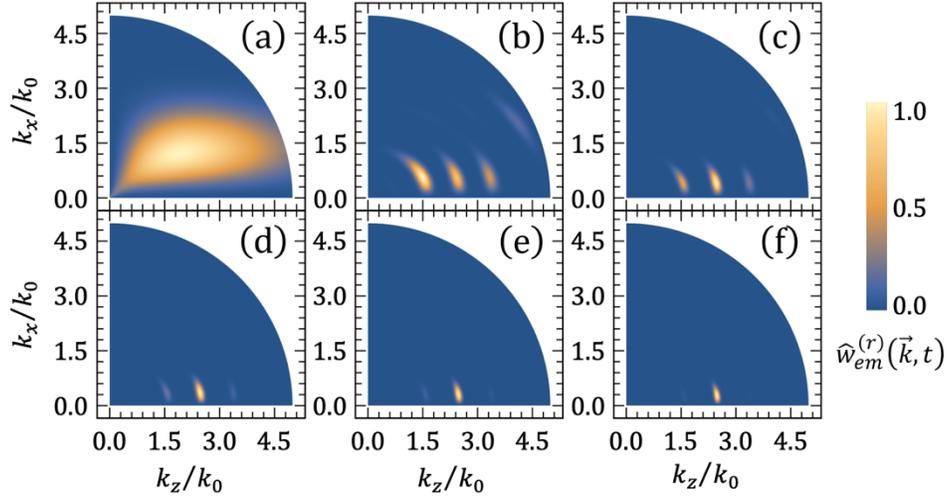

**Fig. 2.** Normalized radiative energy $\widehat{w}_{em}^{(r)}(\vec{k}, t)$ in momentum space at various time instants (a) $t = T^-$, (b) $t = 4T^-$, (c) $t = 7T^-$, (d) $t = 10T^-$, (e) $t = 13T^-$ and (f) $t = 16T^-$ of the APTC in Fig. **1**(a) due to an embedded stationary charge. The initial radiative energy at time $t = 0^-$ is zero, and the other parameters are the same as in Fig. **1**(b-c).

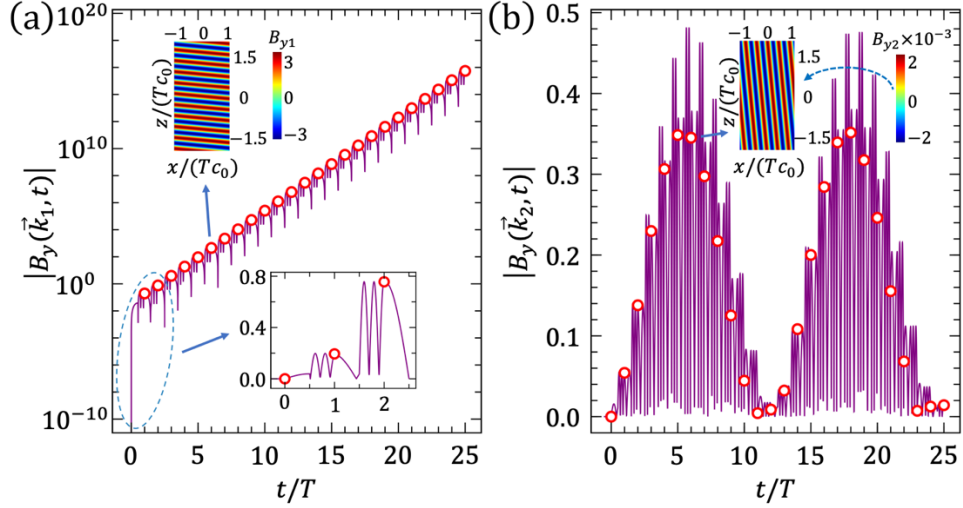

**Fig. 3.** Magnitude of the magnetic-flux intensity component $B_y(\vec{k}, t)$ (in units of $Q\eta_0/k_0$) as a function of time $t$ for the APTC with an embedded stationary charge $Q$, at the momentum $\vec{k}_l$, $l = 1, 2$ of the spherical coordinates (a) $(2.5k_0, 0.094, 0)$ and (b) $(2.5k_0, 1.477, 0)$. The empty circles and the solid lines are theoretical and numerical results respectively. The lower inset in (a) shows the initial stage of the main panel on a linear scale, and the upper insets of both (a) and (b) exhibit the corresponding spatial distribution of the $B_y$ field at time $t = 6T$. The other parameters are the same as Fig. **1**(b-c).

# Supplementary material

# Stationary Charge Radiation in Anisotropic Photonic Time Crystals

Huanan Li,[1] Shixiong Yin,[2,3] Huan He,[1] Jingjun Xu,[1,*] Andrea Alù,[2,3,4,†] and Boris Shapiro[5,‡]

[1]MOE Key Laboratory of Weak-Light Nonlinear Photonics, School of Physics, Nankai University, Tianjin 300071, China

[2]Photonics Initiative, Advanced Science Research Center, City University of New York, New York, New York 10031, USA

[3]Department of Electrical Engineering, City College of The City University of New York, New York, New York 10031, USA

[4]Physics Program, Graduate Center, City University of New York, New York, New York 10016, USA

[5]Department of Physics, Technion-Israel Institute of Technology, Haifa 32000, Israel

## I. Derivations of Eqs. (3) and (6) in the main text

We consider the anisotropic photonic time crystal (APTC) [see Fig. 1(a) in the main text] with a stationary point charge $Q$ at the origin $\vec{r} = 0$, and, in this section, focus on a temporal interface at time $t_s$ when the medium switches from medium 1 (isotropic) to medium 2 (uniaxial crystal). At the temporal interface, the temporal boundary conditions $\vec{D}(\vec{r}, t_s^-) = \vec{D}(\vec{r}, t_s^+)$ and $\frac{\partial \vec{D}}{\partial t}(\vec{r}, t_s^-) =$

* jjxu@nankai.edu.cn
† aalu@gc.cuny.edu
‡ boris@physics.technion.ac.il

$\frac{\partial \vec{D}}{\partial t}(\vec{r}, t_s^+)$ hold for any position $\vec{r}$, leading to the corresponding relationships $\vec{D}(\vec{k}, t_s^-) = \vec{D}(\vec{k}, t_s^+)$ and $\frac{\partial \vec{D}}{\partial t}(\vec{k}, t_s^-) = \frac{\partial \vec{D}}{\partial t}(\vec{k}, t_s^+)$ in momentum space, or equivalently

$$\vec{D}_0^{(1)} + \sum_{n=1}^{2} \vec{D}_n [1, \ 1] \begin{bmatrix} f_n(t_s^-) \\ b_n(t_s^-) \end{bmatrix} = \vec{D}_0^{(2)} + \sum_{n=1}^{2} \vec{D}_n [1, \ 1] \begin{bmatrix} f_n(t_s^+) \\ b_n(t_s^+) \end{bmatrix} \tag{S1a}$$

$$\sum_{n=1}^{2} \vec{D}_n [1, \ 1] \begin{bmatrix} j\omega_n^{(1)} f_n(t_s^-) \\ -j\omega_n^{(1)} b_n(t_s^-) \end{bmatrix} = \sum_{n=1}^{2} \vec{D}_n [1, \ 1] \begin{bmatrix} j\omega_n^{(2)} f_n(t_s^+) \\ -j\omega_n^{(2)} b_n(t_s^+) \end{bmatrix} \tag{S1b}$$

using Eqs. (1) and (5) in the main text, where $\vec{D}_0^{(1)} = \frac{jQ}{k}[\hat{k}_x, \ \hat{k}_y, \ \hat{k}_z]^T$ and $\vec{D}_0^{(2)} = \frac{jQ}{k} \frac{1}{\varepsilon_\perp(\hat{k}_x^2 + \hat{k}_y^2) + \varepsilon_\parallel \hat{k}_z^2} [\varepsilon_\perp \hat{k}_x, \ \varepsilon_\perp \hat{k}_y, \ \varepsilon_\parallel \hat{k}_z]^T$ are the electrostatic solutions in $\vec{k}$-space for the charge $Q$ in medium 1 and 2 respectively, $\vec{D}_1 = [-\hat{k}_y, \hat{k}_x, 0]^T$ and $\vec{D}_2 = [\hat{k}_x \hat{k}_z, \hat{k}_y \hat{k}_z, -(\hat{k}_x^2 + \hat{k}_y^2)]^T$ are the basis states for radiation, and, for brevity, the argument $\vec{k}$ in all quantities has been omitted. Now we intend to project the two vector equations, (S1a) and (S1b), onto the orthogonal triplet consisting of $\vec{k}$, $\vec{D}_1(\vec{k})$ and $\vec{D}_2(\vec{k})$ mentioned in the main text. Since $\vec{D}_0^{(1)} \cdot \vec{k} = \vec{D}_0^{(2)} \cdot \vec{k}$ and $\vec{D}_1 \cdot \vec{k} = \vec{D}_2 \cdot \vec{k} = 0$, the projection of Eqs. (S1a) and (S1b) onto the $\vec{k}$ vector produce trivial identities irrespective of the state $\psi(t) \equiv \begin{bmatrix} \psi_1(t) \\ \psi_2(t) \end{bmatrix}$ with $\psi_n(t) \equiv [f_n(t), \ b_n(t)]^T$. However, based on the projection onto the basis states $\vec{D}_1$ and $\vec{D}_2$, we can obtain

$$\begin{bmatrix} \vec{D}_1 \cdot \vec{D}_0^{(1)} \\ \vec{D}_2 \cdot \vec{D}_0^{(1)} \\ 0 \\ 0 \end{bmatrix} + K^{(1)} \psi(t_s^-) = \begin{bmatrix} \vec{D}_1 \cdot \vec{D}_0^{(2)} \\ \vec{D}_2 \cdot \vec{D}_0^{(2)} \\ 0 \\ 0 \end{bmatrix} + K^{(2)} \psi(t_s^+) \tag{S2}$$

with the coefficient matrices

$$K^{(m)} = \begin{bmatrix} \vec{D}_1 \cdot \vec{D}_1 & \vec{D}_1 \cdot \vec{D}_1 & 0 & 0 \\ 0 & 0 & \vec{D}_2 \cdot \vec{D}_2 & \vec{D}_2 \cdot \vec{D}_2 \\ j\omega_1^{(m)}\vec{D}_1 \cdot \vec{D}_1 & -j\omega_1^{(m)}\vec{D}_1 \cdot \vec{D}_1 & 0 & 0 \\ 0 & 0 & j\omega_2^{(m)}\vec{D}_2 \cdot \vec{D}_2 & -j\omega_2^{(m)}\vec{D}_2 \cdot \vec{D}_2 \end{bmatrix}. \quad (S3)$$

The solution of Eq. (S2) for the state $\psi(t_s^+)$ after switching is simply Eq. (6) in the main text, i.e.,

$$\psi(t_s^+) = S^{(2,1)} + J^{(2,1)}\psi(t_s^-) \quad (S4)$$

with the matching matrix $J^{(2,1)} \equiv [K^{(2)}]^{-1} K^{(1)}$ and the source term $S^{(2,1)} \equiv [K^{(2)}]^{-1} \left[ \vec{D}_1 \cdot \left(\vec{D}_0^{(1)} - \vec{D}_0^{(2)}\right),\ \vec{D}_2 \cdot \left(\vec{D}_0^{(1)} - \vec{D}_0^{(2)}\right),\ 0,\ 0 \right]^T$ given explicitly in the main text. Eq. (3) in the main text can be considered as a special scenario of Eq. (S4) when the charge is absent, i.e., $Q = 0$. In this case, we have $S^{(2,1)} = 0$ and $\psi(t_s^+) = J^{(2,1)}\psi(t_s^-)$, i.e., Eq. (3) in the main text.

## II. Derivation of Eq. (7) in the main text

The conventional temporal transfer matrix formalism [1]-[3] for the radiation field $\psi(t)$ in the APTCs changes dramatically when we introduce the point charge $Q$, since a switching event in the APTCs can create radiation fields by itself in the absence of any initial excitations before the switching, see Eq. (S4). For later purpose, we also write down the matching equation at the other switching event at time $t_s$ when the medium changes from $m = 2$ to $1$ [see Fig. **S1**]

$$\psi(t_s^+) = S^{(1,2)} + J^{(1,2)}\psi(t_s^-) \quad (S5)$$

with the source term $S^{(1,2)} = -[J^{(2,1)}]^{-1} S^{(2,1)}$ and the matching matrix $J^{(1,2)} = [J^{(2,1)}]^{-1}$, which can be derived following a similar line of argument for Eq. (S4). Moreover, as mentioned in the main text, the evolution of the state $\psi(t)$ within each temporal slab is not affected by the charge, and thus determined by the same propagation matrix $F^{(m)}$ [see Eq. (2) in the main text] for the medium $m$. First, we use mathematical induction to prove the following result

$$\psi(n_t T^-) = \frac{\bar{\bar{I}}_4 - M^{n_t}}{\bar{\bar{I}}_4 - M}\left(F^{(1)} - M\right)S^{(1,2)} \tag{S6}$$

for the radiation field $\psi(t)$ after $n_t = 1, 2, 3 \cdots$ modulation cycles without initial excitations, i.e., $\psi(0^-) = 0$. As shown in the main text explicitly, Eq. (S6) is true when $n_t = 1$, i.e., $\psi(T^-) = \left(F^{(1)} - M\right)S^{(1,2)}$. Now we assume that Eq. (S6) is valid for a particular modulation cycle $n_t - 1, n_t \geq 2$, meaning Eq. (S6) for $\psi\{(n_t - 1)T^-\}$ is true:

$$\psi\{(n_t - 1)T^-\} = \frac{\bar{\bar{I}}_4 - M^{n_t-1}}{\bar{\bar{I}}_4 - M}\left(F^{(1)} - M\right)S^{(1,2)}. \tag{S7}$$

It follows that $\psi\{(n_t - 1)T^+\} = S^{(2,1)} + J^{(2,1)}\psi\{(n_t - 1)T^-\}$ using Eq. (S4) for the time interface at time $t_s = (n_t - 1)T$, see Fig. **S1**. This field freely propagate in the medium $m = 2$, for a time interval $T_2$, and we have $\psi\{[(n_t - 1)T + T_2]^-\} = F^{(2)}\psi\{(n_t - 1)T^+\}$. It will be changed to be $\psi\{[(n_t - 1)T + T_2]^+\} = S^{(1,2)} + J^{(1,2)}\psi\{[(n_t - 1)T + T_2]^-\}$ after the second switching event at time $t_s = (n_t - 1)T + T_2$, see Eq. (S5). The additional free propagation in medium 1 of this resulting radiation for duration $T_1$ yields $\psi(n_t T^-) = F^{(1)}\psi\{[(n_t - 1)T + T_2]^+\}$ for the radiation field $\psi(t)$ after $n_t$ modulation cycles. Putting above pieces together, we can produce Eq. (S6) considering the assumption Eq. (S7) together with the relationships $S^{(2,1)} = -J^{(2,1)}S^{(1,2)}$ and $M = F^{(1)}J^{(1,2)}F^{(2)}J^{(2,1)}$. Thus, by mathematical induction given above, Eq. (S6) holds for every modulation cycle $n_t \geq 1$. Finally, noting that $S^{(1,2)} = -s_{21}[0, \ 0, \ 1, \ 1]^T$ and the matrices $F^{(1)} = \text{diag}\{F_1^{(1)}, \ F_2^{(1)}\}$ and $M = \text{diag}\{M_1, M_2\}$ are block diagonal, we obtain from Eq. (S6) $\psi_1(n_t T^-) = 0$ and

$$\psi_2(n_t T^-) = -s_{21} \frac{\bar{\bar{I}}_2 - M_2^{n_t}}{\bar{\bar{I}}_2 - M_2}\left(F_2^{(1)} - M_2\right)\begin{bmatrix}1\\1\end{bmatrix}, \tag{S8}$$

i.e., Eq. (7) in the main text.

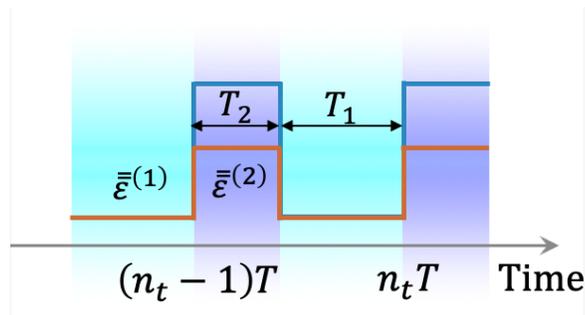

**Fig. S1.** Schematic of part of the APTC [see Fig. 1(a) in the main text] for the derivation of Eq. (S6).

## III. Localization points of radiative energy in momentum space

We first identify the relevant system parameters for controlling the stationary charge's emission features in the APTCs. As shown in the main text, the radiation features are controlled by the band structure of the APTCs for extraordinary waves, which in turn is obtained by solving the eigenvalue problem of the sub-matrix $M_2 = F_2^{(1)}\left(J_2^{(2,1)}\right)^{-1}F_2^{(2)}J_2^{(2,1)}$ of the unit-cell transfer matrix $M$ [see Eq. (4) in the main text]. Below, we rewrite the matching sub-matrix $J_2^{(2,1)}$ and the propagation sub-matrices $F_2^{(1)}$ and $F_2^{(2)}$ [Eqs. (2, 3) in the main text], i.e., the building blocks of $M_2$, explicitly as

$$J_2^{(2,1)} = \frac{1}{2}\begin{bmatrix} \frac{1}{\omega_2^{(2,1)}} + 1 & -\frac{1}{\omega_2^{(2,1)}} + 1 \\ -\frac{1}{\omega_2^{(2,1)}} + 1 & \frac{1}{\omega_2^{(2,1)}} + 1 \end{bmatrix} \tag{S9}$$

and

$$F_2^{(1)} = \mathrm{diag}\{e^{j2\pi\alpha\hat{k}}, \quad e^{-j2\pi\alpha\hat{k}}\} \tag{S10a}$$

$$F_2^{(2)} = \mathrm{diag}\{e^{j2\pi(1-\alpha)\omega_2^{(2,1)}\hat{k}}, \quad e^{-j2\pi(1-\alpha)\omega_2^{(2,1)}\hat{k}}\} \tag{S10b}$$

in terms of the normalized parameters $\omega_2^{(2,1)} \equiv \omega_2^{(2)}/\omega_2^{(1)} = \sqrt{(\hat{k}_x^2 + \hat{k}_y^2)/\hat{\varepsilon}_\parallel + \hat{k}_z^2/\hat{\varepsilon}_\perp}$, $\alpha = T_1/T$ and $\hat{k} = k/(\sqrt{\varepsilon}k_0)$, where $\hat{\varepsilon}_\parallel = \varepsilon_\parallel/\varepsilon$, $\hat{\varepsilon}_\perp = \varepsilon_\perp/\varepsilon$, and $k_0 = 2\pi/Tc_0$. As seen, the band structure of the APTC in momentum space $\vec{k}$ (in units of $\sqrt{\varepsilon}k_0$) for extraordinary waves, dictated by the sub-matrix $M_2$, is determined completely by the three dimensionless system parameters $\hat{\varepsilon}_\parallel$, $\hat{\varepsilon}_\perp$ and $\alpha$. Similarly, we can also show that the above three parameters $\hat{\varepsilon}_\parallel$, $\hat{\varepsilon}_\perp$ and $\alpha$ determine the distribution of the normalized radiative energy density $\widehat{w}_{em}^{(r)}(\vec{k}, t)$ in momentum space $\vec{k}$ (in units of $\sqrt{\varepsilon}k_0$).

For convenience, we can introduce an equivalent set of control parameters for the stationary charge emission in APTCs [see Fig. 1(a) in the main text], which consists of the geometrical average $\bar{\hat{\varepsilon}} \equiv \sqrt{\hat{\varepsilon}_\parallel \hat{\varepsilon}_\perp}$ of the normalized $\hat{\varepsilon}_\parallel$ and $\hat{\varepsilon}_\perp$, and the refractive-index-ratio $\eta \equiv n_o/n_e$ between the ordinary index $n_o = \sqrt{\varepsilon_\perp}$ and the extraordinary index $n_e = \sqrt{\varepsilon_\parallel}$ of the uniaxial crystal, in addition to the aforementioned geometric parameter $\alpha$ of the APTCs. Importantly, the set of parameters $\{\bar{\hat{\varepsilon}}, \eta, \alpha\}$ can control the localization position of $\widehat{w}_{em}^{(r)}(\vec{k}, t)$ in momentum space [see Fig. 2 in the main text], which is directly associated with the maximum of $|\mathrm{Im}(\Omega_{2\sigma}T)|$, i.e., the maximum rate of exponential growth of the gap modes for extraordinary waves residing in the $\vec{k}$-range of initial excitations [4]. We note that the $\vec{k}$-range of initial excitations $\widehat{w}_{em}^{(r)}(\vec{k}, T^-)$ [see, for example, Fig. 2(a) in the main text] is largely tunable via the geometric parameter $\alpha$ of the APTCs. In addition, the radiation of the stationary charge in the APTCs is possible only when the refractive-index-ratio $\eta \neq 1$, and its value can control the anisotropy of the band structure for extraordinary waves, while the refractive-index-average $\bar{\bar{\varepsilon}}$ and the geometric parameter $\alpha$ for the radial distribution of the local peaks of $|\text{Im}(\Omega_{2\sigma}T)|$ control the position of the localization point of $\widehat{w}_{em}^{(r)}(\vec{k}, t)$. For example, when changing the value of the refractive-index-ratio $\eta$ from $\eta > 1$ to $\eta < 1$, the localization position of $\widehat{w}_{em}^{(r)}(\vec{k}, t)$ shifts from closer to the $k_z$-axis [see Fig. 2 in the main text] to close to the $k_x$-axis [see Fig. **S2** below].

## IV. Discussion of experimental aspects

The system parameters used to demonstrate the radiation feature of the stationary charge in the APTCs [see Fig. 2 in the main text] correspond to normalized system parameters $\bar{\bar{\varepsilon}} = 10$, $\eta = 5/2$, and $\alpha = 0.5$. The extreme optical anisotropy here can be exhibited in transition-metal dichalcogenides (TMDCs) such as $NiTe_2$ in the near-infrared and visible ranges [5]-[6]. Furthermore, the optical anisotropy of $NiTe_2$ is largely tunable by controlling the Fermi level via external parameters like electric or magnetic field [5]. Nevertheless, the observation of the radiation feature is constrained by loss and the finite time scales of the switching. Qualitatively, to observe the demonstrated physics, the increment of the amplification due to the modulation in the APTCs must exceed the decrement due to loss, and the material properties need to be switched fast enough to produce time-reflected waves for the emergence of a band structure with open band gaps. In principle, the considerable loss in $NiTe_2$ can be balanced by material gain and the fast modulation is achievable via optical pumping. Currently, a quantitative theory in our scenario dealing with loss and finite switching time involving dispersion doesn't exist yet, which is beyond the scope of the current work and will be explored further in follow-up studies.

We must emphasize that the physics of the stationary charge radiation in APTCs is *not* specific to the extreme parameters used in the main text. In Fig. **S2**, we consider moderate parameters $\{\bar{\bar{\varepsilon}} = 3, \eta = 2/3, \alpha = 0.5\}$ corresponding to $\varepsilon_\perp/\varepsilon = 2, \varepsilon_\parallel/\varepsilon = 4.5$. As shown, the localization phenomenon of radiated energy, associated with the band structure for extraordinary waves [Fig. **S2**(a)], emerges as the number of modulation cycles increases [Fig. **S2**(b)]. We can optimize this set of parameters to further reduce the requirements on the time-switching protocol, for instance further reducing the contrast in switching parameters. Typically this comes at the cost of increasing challenges in localization due to the reduced sharpness of the peak of $|\text{Im}(\Omega_{2\sigma}T)|$ when $\eta \to 1$, and requiring a larger number of modulation cycles.

To implement APTCs experimentally, we expect that metamaterials with switchable, subwavelength and anisotropic inclusions are a viable option. At microwaves, large anisotropy can be introduced by loading a transmission line (TL) with subwavelength unit cells featuring anisotropic electric and/or magnetic responses. For instance, a permeability ratio $|\mu_y/\mu_x| \approx 3.85$ was demonstrated experimentally before at 1.94 GHz [7]. Furthermore, ultrafast time-varying electromagnetic responses in microwaves are more approachable than in optics, especially with recent advances in reconfigurable intelligent surfaces where Arduino and field programmable gate arrays (FPGAs) are widely applied for real-time monitoring, signal processing and voltage control (see, e.g., Ref. [8]). Indeed, time interfaces for electromagnetic waves were recently experimentally realized in a microwave TL metamaterial [9].

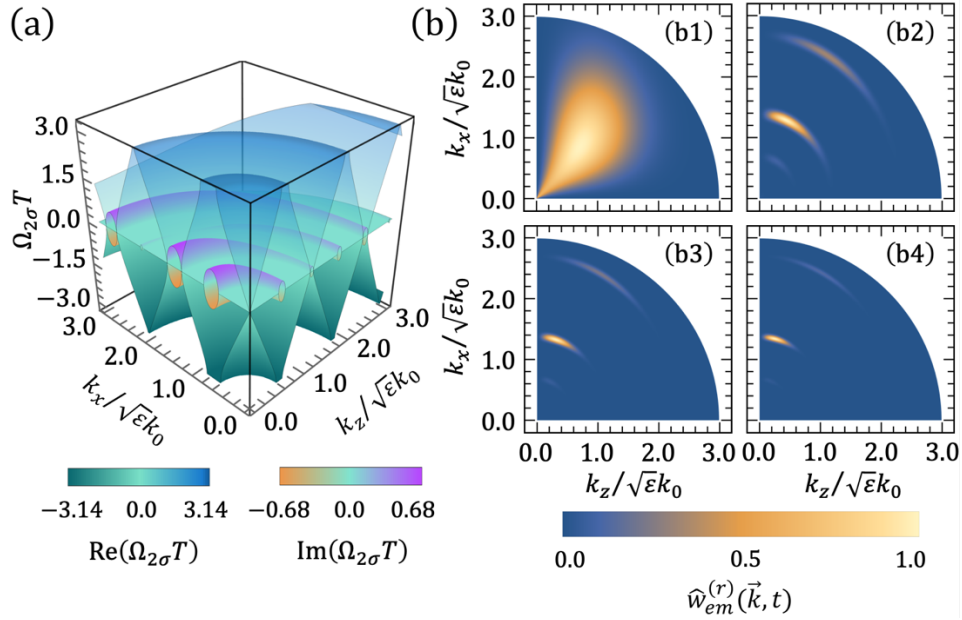

**Fig. S2.** (a) Band structure of the APTC [with $\hat{\varepsilon}_\perp \equiv \varepsilon_\perp/\varepsilon = 2, \hat{\varepsilon}_\parallel \equiv \varepsilon_\parallel/\varepsilon = 4.5$, and $\alpha \equiv T_1/T = 0.5$] in momentum space for extraordinary waves. (b) Normalized radiative energy $\hat{w}_{em}^{(r)}(\vec{k},t)$ in momentum space at various time instants (b1) $t = T^-$, (b2) $t = 7T^-$, (b3) $t = 13T^-$, (b4) $t = 19T^-$ of the APTC in (a) due to an embedded stationary charge in the absence of initial radiation.

By switching on and off the loaded lumped capacitors (82 pF) within ~4 ns, the impedance of the TL was switched between ~25 Ohm and ~50 Ohm, corresponding to a four-fold change in the effective permittivity at the switching time $t_s$, i.e., $\varepsilon(t_s^+)/\varepsilon(t_s^-) \approx 4$. An easy extension to realize APTCs is to replace the isotropic capacitive loads with anisotropic unit cells. Considering the above experimental achievements, the parameters used in Fig. **S2** appear to be readily reachable.

# References

[1] R. Carminal, H. Chen, R. Pierrat, and B. Shapiro, "Universal Statistics of Waves in a Random Time-Varying Medium," Phys. Rev. Lett. **127**, 094101 (2021).

[2] H. Li, S. Yin, E. Galiffi, and A. Alù, "Temporal Parity-Time Symmetry for Extreme Energy Transformations," Phys. Rev. Lett. **127**, 153903 (2021).

[3] J. Xu, W. Mai, and D. H. Werner, "Generalized temporal transfer matrix method: a systematic approach to solving electromagnetic wave scattering in temporally stratified structures," Nanophotonics **11**, 1309 (2022).

[4] In principle, the effects of the APTCs at the $\vec{k}$ values away from the range of initial excitations can be also restricted by material dispersion and thus negligible.

[5] C. Rizza, D. Dutta, B. Ghosh et al., "Extreme Optical Anisotropy in the Type-II Dirac Semimetal $NiTe_2$ for Applications to Nanophotonics," ACS Appl. Nano Mater. **Article ASAP** (2022).

[6] G. A. Ermolaev, D. V. Grudinin, Y. V. Stebunov et al., "Giant optical anisotropy in transition metal dichalcogenides for next-generation photonics," Nat. Commun. **12**, 854 (2021).

[7] J. K. H. Wong, K. G. Balmain, and G. V. Eleftheriades, "Fields in Planar Anisotropic Transmission-Line Metamaterials," IEEE Trans. Antennas Propag. **54**, 2742 (2006).

[8] Q. Ma, G. D. Bai, H. B. Jing, C. Yang, L. Li, and T. J. Cui, "Smart Metasurface with self-adaptively reprogrammable functions," Light Sci. Appl. **8**, 98 (2019).

[9] H. Moussa, G. Xu, S. Yin, E. Galiffi, Y. Radi, and A. Alù, "Observation of Temporal Reflections and Broadband Frequency Translations at Photonics Time-Interfaces," arXiv: 2208.07236 (2022).